\documentclass{article}
\usepackage{amssymb}
\usepackage{amsmath}

\input{tcilatex}
\begin{document}

\title{R{\large adiative Processes in Graphene and Similar Nanostructures at
Strong Electric Fields }}
\author{S.P. Gavrilov\thanks{%
Department of Physics, Tomsk State University, Lenin Prospekt 36, 634050,
Tomsk, Russia; Department of General and Experimental Physics, Herzen State
Pedagogical University of Russia, Moyka emb. 48, 191186 St. Petersburg,
Russia; e-mail: gavrilovsergeyp@yahoo.com; gavrilovsp@herzen.spb.ru} and
D.M. Gitman\thanks{%
Department of Physics, Tomsk State University, Lenin Prospekt 36, 634050,
Tomsk, Russia; P. N. Lebedev Physical Institute, 53 Leninskiy prospect,
119991, Moscow, Russia; Instituto de F\'{\i}sica, Universidade de S\~{a}o
Paulo, CP 66318, CEP 05315-970 S\~{a}o Paulo, SP, Brazil; e-mail:
gitman@if.usp.br}}
\maketitle

\begin{abstract}
Low-energy single-electron dynamics in graphene monolayers and similar
nanostructures is described by the Dirac model, being a 2+1 dimensional
version of massless QED with the speed of light replaced by the Fermi
velocity $v_{F}\simeq c/300$. Methods of strong-field QFT are relevant for
the Dirac model, since any low-frequency electric field requires a
nonperturbative treatment of massless carriers in case it remains unchanged
for a sufficiently long time interval. In this case, the effects of creation
and annihilation of electron-hole pairs produced from vacuum by a slowly
varying and small-gradient electric field are relevant, thereby
substantially affecting the radiation pattern. For this reason, the standard
QED text-book theory of photon emission cannot be of help. We construct the
Fock-space representation of the Dirac model, which takes exact accounts of
the effects of vacuum instability caused by external electric fields, and in
which the interaction between electrons and photons is taken into account
perturbatively, following the general theory (the generalized Furry
representation). We consider the effective theory of photon emission in the
first-order approximation and construct the corresponding total
probabilities, taking into account the unitarity relation.
\end{abstract}

\section{Introduction\label{S1}}

Low-energy single-electron dynamics in graphene monolayers at the charge
neutrality point and similar nanostructures is described by the Dirac model,
being a 2+1 dimensional version of massless QED with the Fermi velocity $%
v_{F}\simeq 10^{6}\mathrm{m/s}$ playing the role of the speed of light in
relativistic particle dynamics. There are actually two species of fermions
in this model, corresponding to excitations about the two distinct Dirac
points in the Brillouin zone of graphene (a distinct pseudo spin is
associated). There also is a (real) spin degeneracy factor $2$. We consider
an infinite flat graphene sample on which a uniform electric field is
applied, directed along the axis $x$ on the plane of the sample. We assume
that the applied field is the $T$-constant electric field that exists during
a macroscopic large time period $T$ comparing to the characteristic time
scale $\Delta t_{\mathrm{st}}=\left( e\left\vert E\right\vert v_{F}/\hbar
\right) ^{-1/2}\gg 0.24\mathrm{fs}$, $10^{-12}\mathrm{s}\gtrsim T>\Delta t_{%
\mathrm{st}}$. This field turns on to $E$ at $-T/2=t_{\mathrm{in}}$ and
turns off to $0$ at $T/2=t_{\mathrm{out}}$.

The electromagnetic field is not confined to the graphene surface, $z=0$,
but rather propagates (with the speed of light $c$) in the ambient $3+1$\
dimensional space-time, where $z$ is the coordinate of axis normal to the
graphene plane. Thus, we have so called reduced QED$_{3,2}$ with distinct
velocities for relativistic dynamics of charged particles and clasical and
quantum electromagnetic fields. Low-frequency ($\omega \lesssim T^{-1}$)
crossed electromagnetic field is radiated in direction orthogonal to
graphene plane by a mean current of pairs created from vacuum, see Ref. \cite%
{GGY12} for details. High-frequency ($\omega \gg T^{-1}$) emission
(absorption) of a photon occures due to a particle state \ transition. E.
g., (1) emission by an electron in initial state, (2) emission with pair
creation from vacuum.

Methods of strong-field QED are relevant for the Dirac model, since any
low-frequency electric field requires a nonperturbative treatment of
massless carriers in case it remains unchanged for a sufficiently long time
interval, $T>\Delta t_{\mathrm{st}}$. In particular, the effect of particle
creation is crucial for understanding the conductivity of graphene,
especially in the so-called nonlinear regime. In this regime, the effects of
creation and annihilation of electron-hole pairs produced from vacuum by a
slowly varying and small-gradient electric field are relevant, thereby
substantially affecting the radiation pattern. For this reason, the standard
QED text-book theory of photon emission (relevant assuming that vacuum is
stable) cannot be of help.

\section{Effective perturbation theory of the photon emission\label{S2}}

We construct the Fock-space representation of the Dirac model, which takes
exact accounts of the effects of vacuum instability caused by external
electric fields, and in which the interaction between electrons and photons
is taken into account perturbatively, following the general theory (the
generalized Furry representation) \cite{FGS}. We use boldface symbols for
three-dimensional vectors and symbols with arrows for in-plane comonents,
for example, $\overrightarrow{r}=\left( x,y\right) $. In the usual dipole
approximation, $z$-dependence of the QED Hamiltonian can be integrated out
and we obtain the Hamiltonian of the electron-photon interaction as%
\begin{eqnarray}
\widehat{\mathcal{H}}_{\mathrm{int}} &\approx &\int \overrightarrow{j}%
_{in}\left( t,\overrightarrow{r}\right) \cdot \left. \overrightarrow{\hat{A}}%
(t,\mathbf{r})\right\vert _{z=0}d\overrightarrow{r},  \notag \\
\overrightarrow{j}_{in}\left( t,\overrightarrow{r}\right) &=&-\frac{ev_{F}}{%
2c}\left[ \hat{\Psi}^{\dag }\left( t,\overrightarrow{r}\right) ,\gamma ^{0}%
\overrightarrow{\gamma }\hat{\Psi}\left( t,\overrightarrow{r}\right) \right]
_{-}\;,  \label{1}
\end{eqnarray}%
where quantum fields $\hat{\Psi}\left( t,\overrightarrow{r}\right) $ and $%
\hat{\Psi}^{\dag }\left( t,\overrightarrow{r}\right) $ obey both the Dirac
equation with the potential $\overrightarrow{A}^{\mathrm{ext}}(t,%
\overrightarrow{r})$ and the standard equal time anticommutation relations.
We decomposed quantum electromagnetic field in the interaction
representation into terms of the annihilation and creation operators of
photons, $C_{\mathbf{k}\vartheta }$ and $C_{\mathbf{k}\vartheta }^{\dagger }$%
: 
\begin{equation}
\mathbf{\hat{A}}(t,\mathbf{r})=c\sum_{\mathbf{k,}\vartheta }\sqrt{\frac{2\pi
\hbar }{\varepsilon V\omega }}\boldsymbol{\epsilon }_{\mathbf{k}\vartheta }%
\left[ C_{\mathbf{k}\vartheta }\,\text{e}^{i(\mathbf{k}\cdot \mathbf{r}%
-\omega t)}+C_{\mathbf{k}\vartheta }^{\dagger }\,\text{e}^{-i(\mathbf{k}%
\cdot \mathbf{r}-\omega t)}\right] \,,  \label{2}
\end{equation}%
where $\vartheta =1,2$ is a polarization index, the $\boldsymbol{\epsilon }_{%
\mathbf{k}\vartheta }$ are unit polarization vectors transversal to each
other and to the wavevector $\mathbf{k}$, $\omega =ck,$ $k=\left\vert 
\mathbf{k}\right\vert $, $V$ is the volume of the box regularization, and $%
\varepsilon $ is the relative permittivity ($\varepsilon =1$ for graphene
suspended in vacuum).

The $\mathrm{in}$- and $\mathrm{out}$- operators of creation and
annihilation of electrons ($a_{n}^{\dag }$, $a_{n}$) and holes ($b_{n}^{\dag
}$, $b_{n}$) \ are defined by the two representations of the quantum Dirac
field $\hat{\Psi}\left( t,\overrightarrow{r}\right) $ as 
\begin{align}
\hat{\Psi}\left( t,\overrightarrow{r}\right) & =\sum_{n}\left[ a_{n}(\mathrm{%
in})\;_{+}\psi _{n}\left( t,\overrightarrow{r}\right) +b_{n}^{\dagger }(%
\mathrm{in})\;_{-}\psi _{n}\left( t,\overrightarrow{r}\right) \right]  \notag
\\
& =\sum_{n}\left[ a_{n}(\mathrm{out})\;^{+}\psi _{n}\left( t,\overrightarrow{%
r}\right) +b_{n}^{\dagger }(\mathrm{out}))\;^{-}\psi _{n}\left( t,%
\overrightarrow{r}\right) \right] ,  \label{f6}
\end{align}%
where $_{\zeta }\psi _{n}\left( t,\overrightarrow{r}\right) $ and $^{\zeta
}\psi _{n}\left( t,\overrightarrow{r}\right) $ are $\mathrm{in}$- and $%
\mathrm{out}$-solutions of the Dirac equation with the potential $%
\overrightarrow{A}^{\mathrm{ext}}(t,\overrightarrow{r})$ for given quantum
numbers $n$ and well-defined sign of frequency $\zeta $ either before
turning on or after turning off of a field, respectively. They related by a
linear transformation of the form: 
\begin{eqnarray}
^{\zeta }\psi _{n}\left( t,\overrightarrow{r}\right) &=&g_{n}(_{+}\mid
^{\zeta })\,_{+}\psi _{n}\left( t,\overrightarrow{r}\right) +g_{n}(_{-}\mid
^{\zeta })\,_{-}\psi _{n}\left( t,\overrightarrow{r}\right) \,,  \notag \\
_{\zeta }\psi _{n}\left( t,\overrightarrow{r}\right) &=&g_{n}\left(
^{+}|_{\zeta }\right) \,^{+}\psi _{n}\left( t,\overrightarrow{r}\right)
+g_{n}\left( ^{-}|_{\zeta }\right) \,^{-}\psi _{n}\left( t,\overrightarrow{r}%
\right) \,,  \label{f7}
\end{eqnarray}%
where the $g^{\prime }$s are some complex coefficients. Here the notation $%
g\left( ^{\zeta ^{\prime }}|_{\zeta }\right) =g\left( _{\zeta }|^{\zeta
^{\prime }}\right) ^{\ast }$ is used. These coefficients obey the unitarity
relations which follow from the orthonormalization and completness relations
for the corresponding solutions. It is known that all $g^{\prime }$s can be
expressed in terms of two of them, e.g. of $g\left( _{+}\left\vert
^{+}\right. \right) $ and $g\left( _{-}\left\vert ^{+}\right. \right) $.
However, even the latter coefficients are not completely independent, 
\begin{equation}
\left\vert g_{n}\left( _{-}\left\vert ^{+}\right. \right) \right\vert
^{2}+\left\vert g_{n}\left( _{+}\left\vert ^{+}\right. \right) \right\vert
^{2}=1.  \label{f9}
\end{equation}%
Then a linear canonical transformation (Bogolyubov transformation) between $%
\mathrm{in}$- and $\mathrm{out}$- operators which follows from Eq.~(\ref{f6}%
) is defined by these coefficients.

The initial and final states with definite numbers of charged particles and
photons can be generally written in the following way:

\begin{eqnarray*}
|\mathrm{in} &>&=C^{\dagger }\ldots b^{\dagger }\left( \mathrm{in}\right)
\ldots a^{\dagger }\left( \mathrm{in}\right) \ldots |0,\mathrm{in}\rangle ,
\\
|\mathrm{out} &>&=C^{\dagger }\ldots b^{\dagger }\left( \mathrm{out}\right)
\ldots a^{\dagger }\left( \mathrm{out}\right) \ldots |0,\mathrm{out}\rangle .
\end{eqnarray*}%
\ The $\mathcal{S}$-matrix or the scattering operator in the first-order
approximation with respect of electron-photon interaction (it is exact with
respect of an interaction with an external field) is 
\begin{equation}
\mathcal{S}\approx 1+i\Upsilon ^{\left( 1\right) },\;\;\Upsilon ^{\left(
1\right) }=-\frac{1}{\hbar }\int_{-\infty }^{\infty }\widehat{\mathcal{H}}_{%
\mathrm{int}}dt  \label{6}
\end{equation}%
In general, the emission of a single photon by an electron is accompanied by
the creation of $M\geq 0$ electron-hole pairs from the vacuum by the
quasiconstant electric field: 
\begin{eqnarray}
\mathcal{P}_{M}\left( \left. \mathbf{k}\vartheta \right\vert \overset{+}{l}%
\right) &=&\sum_{\{m\}\left\{ n\right\} }\left[ M!\left( M+1\right) !\right]
^{-1}\left\vert \left\langle 0,\mathrm{out}\right\vert b_{n_{M}}\left( 
\mathrm{out}\right) \ldots b_{n_{1}}\left( \mathrm{out}\right) \right. 
\notag \\
&&\times \left. a_{m_{M+1}}\left( \mathrm{out}\right) \ldots a_{m_{1}}\left( 
\mathrm{out}\right) C_{\mathbf{k}\vartheta }i\Upsilon ^{\left( 1\right)
}a_{l}^{\dagger }(\mathrm{in})|0,\mathrm{in}\rangle \right\vert ^{2}.
\label{7}
\end{eqnarray}%
\ The probability of transition from the single-electron state characterized
by the quantum numbers $l$ with the emission of one photon with given $%
\mathbf{k}$, $\vartheta $ and the production of arbitrary number of pairs
from the vacuum, that is, the total probability of the emission of the given
photon from the single-electron state, is%
\begin{equation}
\mathcal{P}\left( \left. \mathbf{k}\vartheta \right\vert \overset{+}{l}%
\right) =\sum_{M=0}^{\infty }\mathcal{P}_{M}\left( \left. \mathbf{k}%
\vartheta \right\vert \overset{+}{l}\right) .  \label{8}
\end{equation}%
The probability of the process with the emission of one photon with given $%
\mathbf{k}$, $\vartheta $ and the production of $M\geq 1$ arbitrary pairs
from the vacuum is%
\begin{eqnarray}
\mathcal{P}_{M}\left( \mathbf{k,}\vartheta \right) &=&\sum_{\{m\}\left\{
n\right\} }\left( M!\right) ^{-2}\left\vert \left\langle 0,\mathrm{out}%
\right\vert b_{n_{M}}\left( \mathrm{out}\right) \ldots b_{n_{1}}\left( 
\mathrm{out}\right) \right.  \label{9} \\
&&\times \left. a_{m_{M}}\left( \mathrm{out}\right) \ldots a_{m_{1}}\left( 
\mathrm{out}\right) c_{\mathbf{k}\vartheta }i\Upsilon ^{\left( 1\right) }|0,%
\mathrm{in}\rangle \right\vert ^{2}.  \notag
\end{eqnarray}%
The total probability of the emission of the given photon from the vacuum
and the production of an arbitrary number of pairs from the vacuum is%
\begin{equation}
\mathcal{P}\left( \mathbf{k,}\vartheta \right) =\sum_{M=1}^{\infty }\mathcal{%
P}_{M}\left( \mathbf{k,}\vartheta \right) .  \label{10}
\end{equation}

\ \ The unitary transformation $V$ relates the \textrm{in} and \textrm{out}%
\textbf{- }Fock spaces, $\;|\mathrm{in}\rangle =V|\mathrm{out}\rangle $. It
means that we can pass from the basis of the final Fock space to the basis
of the initial Fock space and, for example, represent the total probabilty (%
\ref{8}) as%
\begin{eqnarray}
\mathcal{P}\left( \left. \mathbf{k}\vartheta \right\vert \overset{+}{l}%
\right) &=&\sum_{n}\left\vert w_{in}^{\left( 1\right) }\left( \overset{+}{n}%
;\left. \mathbf{k}\vartheta \right\vert \overset{+}{l}\right) \right\vert
^{2},  \notag \\
w_{in}^{\left( 1\right) }\left( \overset{+}{n};\left. \mathbf{k}\vartheta
\right\vert \overset{+}{l}\right) &=&\left\langle 0,\mathrm{in}\right\vert
a_{n}\left( \mathrm{in}\right) C_{\mathbf{k}\vartheta }i\Upsilon ^{\left(
1\right) }a_{l}^{\dagger }(\mathrm{in})|0,\mathrm{in}\rangle .  \label{11}
\end{eqnarray}%
Note that if the number of pair created is not small then the matrix element 
$w_{in}^{\left( 1\right) }\left( \overset{+}{n};\left. \mathbf{k}\vartheta
\right\vert \overset{+}{l}\right) $ is quite distinct from the amlitude of
the relative probability for a one-particle transition with the emission of
a photon,%
\begin{equation*}
w^{\left( 1\right) }\left( \overset{+}{n};\left. \mathbf{k}\vartheta
\right\vert \overset{+}{l}\right) =\frac{\left\langle 0,\mathrm{out}%
\right\vert a_{n}\left( \mathrm{out}\right) C_{\mathbf{k}\vartheta
}i\Upsilon ^{\left( 1\right) }a_{l}^{\dagger }(\mathrm{in})|0,\mathrm{in}%
\rangle }{\langle 0,\mathrm{out}|0,\mathrm{in}\rangle }.
\end{equation*}

\section{Characteristics for the emission of a photon by an electron\label%
{S3}}

We apply this theory to the calculation of total probability for emission of
a photon by an electron in a constant electric field. We defined an
orthonormal triple%
\begin{eqnarray}
\mathbf{k/}k &=&(\sin \theta \cos \phi ,\,\sin \theta \sin \phi ,\,\cos
\theta )\,,  \notag \\
\boldsymbol{\epsilon }_{\mathbf{k}1} &=&\mathbf{e}_{z}\times \mathbf{k/}%
\left\vert \mathbf{e}_{z}\times \mathbf{k}\right\vert ,\quad \boldsymbol{%
\epsilon }_{\mathbf{k}2}=\mathbf{k\times }\boldsymbol{\epsilon }_{\mathbf{k}%
1}/\left\vert \mathbf{k\times }\boldsymbol{\epsilon }_{\mathbf{k}%
1}\right\vert  \label{12}
\end{eqnarray}%
then 
\begin{eqnarray*}
\boldsymbol{\epsilon }_{\mathbf{k}1} &=&(-\sin \phi ,\,\cos \phi ,\,0)\,, \\
\boldsymbol{\epsilon }_{\mathbf{k}2} &=&(-\cos \theta \cos \phi ,\,-\cos
\theta \sin \phi ,\,\sin \theta )\,
\end{eqnarray*}%
for $\mathbf{k}$ in the upper spatial region, $k_{z}\geq 0$. Using the
parametrization , $d\mathbf{k=}c^{-3}\omega ^{2}d\omega d\Omega $, we find
that the probabilty of the emission per unit frequency and solid angle $%
d\Omega $ is%
\begin{eqnarray}
\frac{d\mathcal{P}\left( \left. \mathbf{k}\vartheta \right\vert \overset{+}{%
\overrightarrow{p}}\right) }{d\omega d\Omega } &=&\frac{\alpha }{\varepsilon 
}\left( \frac{v_{F}}{c}\right) ^{2}\frac{\omega \Delta t_{st}^{2}}{\left(
2\pi \right) ^{2}}\left. \left\vert M_{\overrightarrow{p}^{\prime }%
\overrightarrow{p}}^{+}\right\vert ^{2}\right\vert _{\overrightarrow{p}%
^{\prime }=\overrightarrow{p}-\hbar \vec{k}}\;,  \notag \\
M_{\overrightarrow{p}^{\prime }\overrightarrow{p}}^{+}
&=&v_{F}^{2}SC^{\prime }C\exp \left( -i\omega \frac{p_{x}+p_{x}^{\prime }}{%
2eE}\right)  \notag \\
&&\times \left[ (1-i)\sqrt{\frac{eE\hbar }{v_{F}}}\zeta p_{y}^{\prime }\chi
_{\vartheta }^{1,0}Y_{10}+(1+i)\sqrt{\frac{eE\hbar }{v_{F}}}\zeta p_{y}\chi
_{\vartheta }^{0,1}Y_{01}\right.  \notag \\
&&+\left. p_{y}^{\prime }p_{y}\chi _{\vartheta }^{1,1}Y_{11}+2\frac{eE\hbar 
}{v_{F}}\chi _{\vartheta }^{0,0}Y_{00}\right] ,  \notag \\
\;\;C &=&\left( 2eE\hbar v_{F}S\right) ^{-1/2}\exp \left( -\frac{\pi \lambda 
}{8}\right) ,\;\lambda =\frac{v_{F}p_{y}^{2}}{eE\hbar },\;C^{\prime }=\left.
C\right\vert _{p_{y}\rightarrow p_{y}^{\prime }}\;,  \label{13}
\end{eqnarray}%
where $\alpha =e^{2}/c\hslash $ is the fine structure constant, $S$ is the
graphene area, $\chi _{\vartheta }^{\left( 1+s^{\prime }\right) /2,\left(
1+s\right) /2}=U_{s^{\prime }}^{\dagger }\gamma ^{0}\overrightarrow{\gamma }%
\cdot \vec{\epsilon}_{\mathbf{k}\vartheta }U_{s}\,$\ and%
\begin{align*}
\chi _{1}^{0,0}& =-\chi _{1}^{1,1}=\sin \phi ,\;\chi _{1}^{1,0}=-\chi
_{1}^{0,1}=i\zeta \cos \phi ; \\
\chi _{2}^{0,0}& =-\chi _{2}^{1,1}=\cos \theta \cos \phi ,\;\chi
_{2}^{1,0}=-\chi _{2}^{0,1}=-i\zeta \cos \theta \sin \phi .
\end{align*}%
Here $Y_{j^{\prime }j}\left( \rho \right) $ is the Fourier transformation of
the product of the Weber parabolic cylinder functions, 
\begin{eqnarray}
Y_{j^{\prime }j}\left( \rho \right) &\simeq &\int_{-\infty }^{+\infty
}D_{-\nu ^{\prime }-j^{\prime }}[-(1+i)u]D_{\nu -j}[-(1-i)u]\text{e}^{i\rho
u}du,  \notag \\
\quad \nu &=&\frac{i\lambda }{2}\,,\;\nu ^{\prime }=\frac{i\lambda ^{\prime }%
}{2}\,,\;\lambda ^{\prime }=\left. \lambda \right\vert _{p_{y}\rightarrow
p_{y}^{\prime }}\;,\quad \rho \approx \Delta t_{st}\omega .  \label{14}
\end{eqnarray}

\ Applying the saddle-point method to the integral (\ref{14}), we establish
the law of conservation of a kinetic energy, 
\begin{equation}
v_{F}\left( 2eEt+p_{x}+p_{x}^{\prime }\right) =\hbar \omega ,  \label{15}
\end{equation}%
at the saddle-point, $u=\rho /2$. The wide high frequency range follows as%
\begin{equation}
2\Delta t_{st}^{-1}<\omega <2\Delta t_{st}^{-2}T,\;\;t_{st}^{-1}T\gg 1.
\label{16}
\end{equation}%
We find the formation interval for an emission a photon with given $\mathbf{k%
}$. The center of formation interval for given initial momentum $p_{x}$ is 
\begin{equation}
t_{c}=-\frac{p_{x}}{eE}+\frac{\omega \Delta t_{st}^{2}}{2}.  \label{17}
\end{equation}%
The width of the formation interval $\Delta t$ is determinated by an
electric field only:%
\begin{equation}
\Delta t\sim \Delta t_{\mathrm{st}}=\left( \left\vert eE\right\vert
v_{F}/\hbar \right) ^{-1/2}\approx \frac{2.6}{\sqrt{a}}\times 10^{-14}%
\mathrm{s},  \label{18}
\end{equation}%
where 
\begin{equation*}
E=aE_{0},\;\;E_{0}=1\times 10^{6}\mathrm{V/m},\;\;7\times 10^{-4}\ll a\ll 8.
\end{equation*}%
It can be shown that leading contribution to the probability (\ref{13}) is
from terms with $Y_{00}$ and $Y_{01}$.

Taking into account that $\left\vert \lambda -\lambda ^{\prime }\right\vert
\lesssim 1$, we find the main contribution to Eq.~(\ref{13}) as%
\begin{eqnarray}
\left\vert M_{\overrightarrow{p}^{\prime }\overrightarrow{p}}^{+}\right\vert
^{2} &\approx &\left( 2\pi \right) ^{2}f\left( \lambda \right) e^{-3\pi
\lambda ^{\prime }/4}\left\vert \chi _{\vartheta }^{0,1}\right\vert ^{2}, 
\notag \\
f\left( \lambda \right) &=&\frac{\sinh \left( \pi \lambda /2\right) }{2\pi %
\left[ \left( \lambda /2\right) ^{2}+1\right] }e^{-\pi \lambda /4}
\label{19}
\end{eqnarray}%
at$\sqrt{\lambda }\sim 1$. This is Gaussian function of $k$ at fixed $\theta
\neq 0$\ and $\phi \neq 0$, where $p_{y}/\hbar $\ is the position of the
center of the peak. We see polarized emission to directions $\phi
\rightarrow 0$ ($k_{y}\rightarrow 0$) and $\phi \rightarrow \pm \pi /2$ ($%
k_{x},k_{z}\rightarrow 0$). The probabilty of unpolarized emission per unit
frequency and solid angle is%
\begin{equation}
\sum_{\vartheta =1,2}\frac{d\mathcal{P}\left( \left. \mathbf{k}\vartheta
\right\vert \overset{\pm }{\overrightarrow{p}}\right) }{d\omega d\Omega }=%
\frac{\alpha }{\varepsilon }\left( \frac{v_{F}}{c}\right) ^{2}\omega \Delta
t_{st}^{2}f\left( \lambda \right) e^{-3\pi \lambda ^{\prime }/4}\left( 1-%
\frac{k_{y}^{2}}{k^{2}}\right) .  \label{20}
\end{equation}%
For any given $p_{y}$ and $k_{y}/k$, the maximum probabilty is realised with 
$\lambda ^{\prime }\rightarrow 0$ ($k_{y}\sim p_{y}/\hbar $). The angular
distribution is maximal at $k_{y}\rightarrow 0$ ($\mathbf{k}$ is in plane
that is orthogonal to graphene and parallel with an electric field $\mathbf{E%
}$.

We suggest the emission of a photon by an electron in graphene in the
presence of a constant electric field for experimental observations.

\section*{Acknowledgements}

S. P. G. and D.M. G. were supported by a grant from the Russian Science
Foundation, Research Project No. 15-12-10009.

\end{document}